# Capturing long-range interaction with reciprocal space neural network

Hongyu Yu,[a,b,*], Liangliang Hong[a,b,*], Shiyou Chen,[b,c] Xingao Gong,[a,b] Hongjun Xiang[a,b,†]

[a]*Key Laboratory of Computational Physical Sciences (Ministry of Education), Institute of Computational Physical Sciences, and Department of Physics, Fudan University, Shanghai 200433, China*
[b]*Shanghai Qi Zhi Institute, Shanghai 200030, China*
[c]*State Key Laboratory of ASIC and System, School of Microelectronics, Fudan University, Shanghai 200433, China*

**Abstract**:

Machine Learning (ML) interatomic models and potentials have been widely employed in simulations of materials. Long-range interactions often dominate in some ionic systems whose dynamics behavior is significantly influenced. However, the long-range effect such as Coulomb and Van der Wales potential is not considered in most ML interatomic potentials. To address this issue, we put forward a method that can take long-range effects into account for most ML local interatomic models with the reciprocal space neural network. The structure information in real space is firstly transformed into reciprocal space and then encoded into a reciprocal space potential or a global descriptor with full atomic interactions. The reciprocal space potential and descriptor keep full invariance of Euclidean symmetry and choice of the cell. Benefiting from the reciprocal-space information, ML interatomic models can be extended to describe the long-range potential including not only Coulomb but any other long-range interaction. A model NaCl system considering Coulomb interaction and the $Ga_xN_y$ system with defects are applied to illustrate the advantage of our approach. At the same time, our approach helps to improve the prediction accuracy of some global properties such as the band gap where the full atomic interaction beyond local atomic environments plays a very important role.  In summary, our work has expanded the ability of current ML interatomic models and potentials when dealing with the long-range effect, hence paving a new way for accurate prediction of global properties and large-scale dynamic simulations of systems with defects.

**Main text**

*Introduction.* Nowadays, molecular dynamics simulations and the Monte Carlo method are widely used to study the properties of molecules or materials from a microscopic perspective. To perform these numerical methods, a reliable mapping function from the atomic configuration of the system to its energy (or forces on each atom) is necessarily required. In general, there are two ways to construct the mapping function, one taking the electronic structure of the system into account, while the other only considers the atomic elements and positions as the independent variables. A typical representative of the former is the density-functional theory (DFT), which tends to give a high-precision result at the cost of computational efficiency. The latter, often named atomistic potentials, has not only a long history of development but also a rather large and complex family. With the emergence of artificial intelligence, machine learning methods are generally believed to combine the advantages of the above two schemes in terms of both accuracy and computational efficiency. Therefore, more and more researchers have tried to develop new machine-learning potential models and a series of remarkable achievements have been made.

So far, many types of ML potentials, such as high-dimensional neural network potentials (HDNNPs) [1], Gaussian approximation potentials (GAPs) [2], kernel ridge regression methods (KRR) [3], moment tensor potentials (MTPs) [4], spectral neighbor analysis potentials (SNAPs) [5] and message-passing neural networks (MPNN) [6], have gained much attention. Behler and his coworkers proposed a classification scheme [7], which roughly classifies all the ML models into four generations based on whether the short-range interaction, long-range interaction, and non-local effects can be properly considered in the model. The first generation of NNPs was introduced in the 1990s [8]. They are only applicable to systems consisting of a few atoms due to their simple architecture. The second-generation ML potentials, including but not limited to GAPs, MTPs, SNAPs, and HDNNPs, have greatly overcome the shortcomings of the previous generation and became the most widely used models nowadays. The significant improvement of this generation mainly owes to three key concepts: nearsightedness, invariant descriptors, and active learning. Nearsightedness means the total energy of a system can be expressed as a summation of atomic contributions which depend on the local chemical environment inside a cutoff radius. As the price paid for nearsightedness, the second-generation ML potentials tend to neglect all the long-range interactions, which limits their accuracy in some ionic systems.

However, the long-range interaction cannot be neglected when the potential decays algebraically with a power smaller than the spatial dimension and influences the dynamic behavior of the system [9,10]. To solve this issue, ML potentials of the third generation have been developed in which the contribution of electrostatic interactions is calculated analytically with the atomic partial charges predicted by the second set of neural networks [11]. As a further improvement, the fourth-generation ML potentials incorporate a global charge equilibration scheme into the previous framework, thus being applicable to systems with non-local charge transfer or different charge states [12]. There have been some ML potentials attempting to capture long-range interaction [11–15]. Most of them are focused on the Coulomb interaction and neglect other types of long-range interaction such as the Van der Waals interaction. Therefore, they are only capable of learning the Coulomb interaction, which limits their expressing ability to some extent. Meanwhile, it's usually hard to determine the quantity of the charge with high accuracy leading to an avoidless deviation. For molecules without periodic boundary conditions (PBC), the long-range interaction can be learned within the framework of the transformer as a full connection atomic graph [16–18]. Unfortunately, the powerful transformer can not be directly applied to a crystal system with PBC due to the requirement of invariance of the choice of the cell. The crystal graph is usually built with a local connection between atoms based on the interatomic distance and a full atomic connection cannot be built considering the PBC. As a result, the fully connected graph transformer has not been employed in the PBC crystal yet and can't be used to learn the long-range interaction. While MPNN can extend the range of interactions considered to a longer distance thanks to the message-passing mechanism [19–23], it learns a semi-long-range interaction depending on the layer number and suffers from the difficulty of parallelism in the simulation of large systems.

Motivated by the property of the Fourier transform, we believe the long-range interaction can be captured in the reciprocal space for the system with PBC. In solid-state physics, the reciprocal space plays a fundamental role and is widely employed to investigate the property of crystals with PBC, particularly in the theory of diffraction and electron structure with Bloch's theorem [24,25]. To encode the structure, the relative positions between neighbor atoms in real space are usually used while the atomic information in reciprocal space is barely employed [25]. In previous work, the diffraction fingerprint descriptor in reciprocal space is built and the 2D diffraction fingerprint descriptor is designed by Ziletti *et al* [26]. It's constructed by simulating the scattering of an incident plane wave through the crystal as a 2D image with the invariance of system size, atomic permutations, and the number of atoms but without invariance of rotation. Also, Wang *et al* built the model with a reciprocal space representation from elastic x-ray scattering theory without the information on the chemical compositions. Meanwhile, Ren *et al.* proposed the combined use of real-space and reciprocal-space descriptors for the inverse design of crystals though the reciprocal-space descriptor is not invariant with different choices of the cell [27]. An ideal reciprocal descriptor with full atomic information and invariance of E(3) symmetry and choice of the cell is demanding for a better description of the crystal structure.

In this work, we put forward the reciprocal space potential and the reciprocal space descriptor that is invariant of E(3) symmetry and choice of the cell, which helps to capture the long-range interatomic potential and the full atomic interaction information for a better prediction of the global properties.

***Method: Long-range Potential in Reciprocal Space***. In this work, we aim to build an approach that can describe the crystal with full atomic interaction including both the local and long-range contributions considering periodic boundary conditions (PBC) shown in Figure 1. We focus on the long-range contributions part as the local atomic contributions can be built by any local descriptors or crystal graph neural network (CGNN) from the real space of crystal with a cutoff distance as SOAP and Dimenet [23,28–40]. For the long-range potential, we start from the Ewald summation which is commonly used for computing the Coulomb interaction interactions in periodic systems. The Ewald summation divides the Coulomb interaction into contributions of short-range and long-range. The short-range part of the Ewald summation can be easily included by current models in the real space with a cutoff so we are more interested in the long-range part of the Ewald summation. The long-range part is calculated in reciprocal space with the generalized formalism $E_{lr} = \frac{1}{\Omega}\sum_k \Phi(|\vec{k}|)|\rho(\vec{k},\vec{r})|^2$, $\rho(\vec{k},\vec{r}) = \exp(-b^2)$, $S(\vec{k},\vec{r}) = \sum_j q_j \exp(-i\vec{k}\cdot\vec{r})$, $b = \frac{|\vec{k}|}{2\alpha}$ with a cutoff distance for integer k-vector in reciprocal space [41–43]. For the long-range potential in the form of $\frac{1}{r^n}$, $\Phi_n(|\vec{k}|) = \pi^{n-\frac{3}{2}}|\vec{k}|^{(n-3)}\Gamma\left(-\frac{n}{2}+\frac{3}{2},b\right)$ with $\Gamma$ as the incomplete gamma function. For Coulomb interaction, $\Phi_1(|\vec{k}|) = \frac{\exp(-b^2)}{2\pi|\vec{k}|^2}$. For Van der Waals interaction, $\Phi_6(|\vec{k}|) = \frac{\pi^{\frac{9}{2}}}{3}|\boldsymbol{k}|^3[\pi^{\frac{1}{2}}\text{ERFC}(b) + \left(\frac{1}{2b^3}-\frac{1}{b}\right)\exp(-b^2)]$ with ERFC as the complementary error function [44]. Based on the general formalism $E_{lr}$ from the Ewald summation, we build our reciprocal space potential beyond an explicit form such as Coulomb potential. With fully connected networks (FCN), it is constructed as $E_{long} = \frac{1}{\Omega}\sum_k \text{FCN}(|\vec{k}|)|S(\vec{k},\vec{r})|^2$ and the structural factor $S(\vec{k},\vec{r}) = \sum_j q_j \exp(-i\vec{k}\cdot\vec{r})$ with the charge of the atom as $q_j$. A cutoff length $k_{cutoff}$ of $\vec{k}$ is manully set and all the integer reciprocal space vector $\vec{k}$ within the length $k_{cutoff}$ is sampled as the Ewald summation. $\Phi(|\vec{k}|)$ in Ewald summation is learned by a neural network consisting of multilayer perceptrons and residual layers shown in Figure 2. As a result, any formalism of long-range interaction can be considered in our approaches with an implicit form. As for the atomic charge $q_j$, basically, we use an embedding layer of species to represent as $q_j = \text{Embedding}(z_j) - q_{neutral}$ and $q_{neutral}$ is the average charge of the entire system which is subtracted based on the neutral charge assumption. For complex systems where we should consider the change of the atoms, we employ the crystal graph neural network to further predict the charge as $q_j = \text{Embedding}(z_j) + h_j^{GNN} - q_{neutral}$ with a fixed charge based on the local atomic environment and neutral charge adjustment. If non-local charge transfer should be included, the electronegativities of atoms are first learned as $\chi_j = \text{Embedding}(z_j) +$ $h_j^{GNN}$ and after charge equilibration, atomic charges $q_j$ are predicted as the fourth-generation NN potential [12]. Information on full atomic interaction is included in $S(\vec{k},\vec{r})$. In a word, any form of long-range interaction with or without considering charge fluctuation and transfer can be learned within our reciprocal space potential framework.

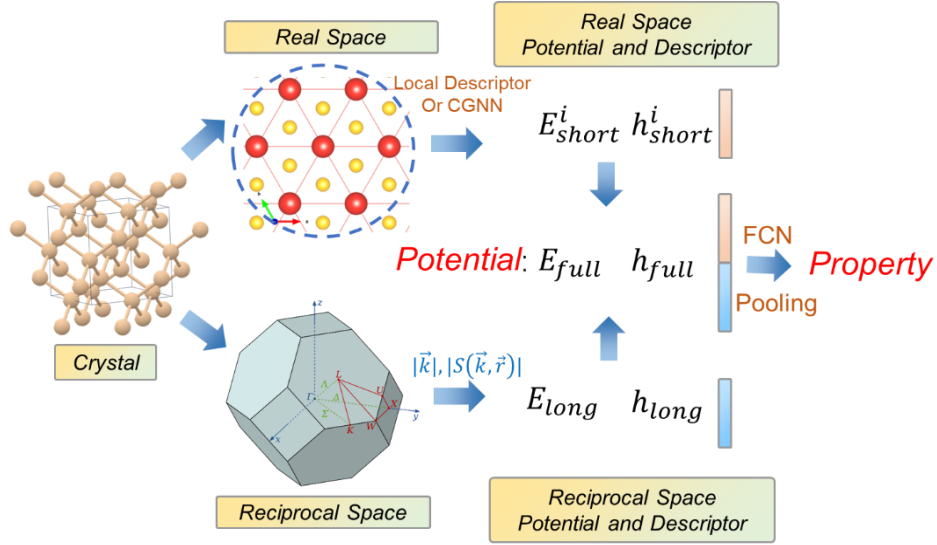

**Figure 1.** Illustration of the construction of crystal features $h_{full}$ including reciprocal space features and original real space features and $E_{full}$ with local energy and reciprocal space energy. The long-range interaction of arbitrary form is learned in reciprocal space.

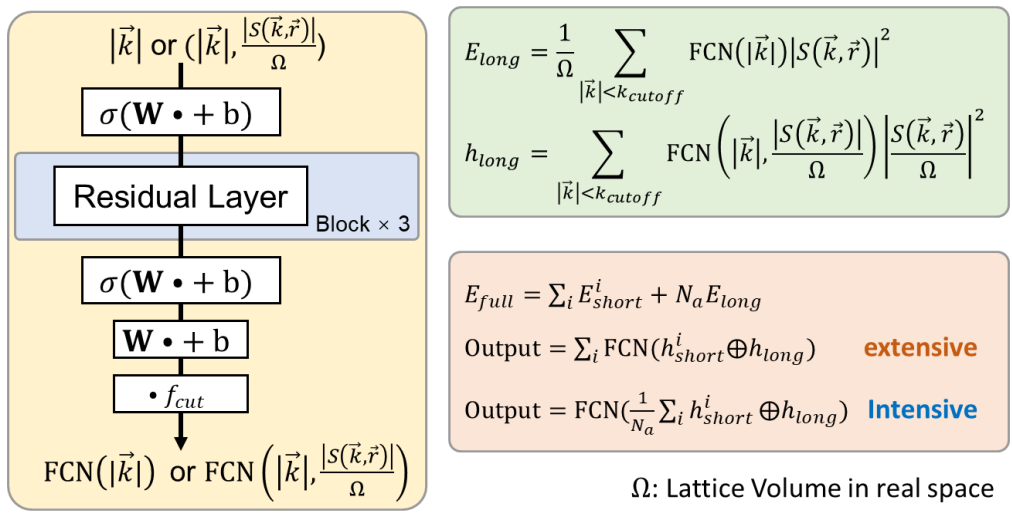

**Figure 2.** The details of the construction of the long-range potential and descriptor.

Similar to the usual local energy methods, our reciprocal space potential keeps invariant under operations of E(3) symmetry including rotation, translation, and inversion with different choices of the cell. As for rotation operation, the length of the sampled $\vec{k}$ vectors keep invariant and $(R\vec{k}) \cdot (R\vec{r}) = \vec{k} \cdot \vec{r}$. As for translation, it only affects the phase of structure factor $S(\vec{k}, \vec{r})$ and $|S(\vec{k}, \vec{r})|^2$ keeps invariant as $S(\vec{k}, \vec{r} + \vec{T}) = \sum_j q_j \exp(-i\vec{k} \cdot \vec{r} - i\vec{k} \cdot \vec{T}) = \exp(-i\vec{k} \cdot \vec{T}) S(\vec{k}, \vec{r})$. As for inversion, it only reverses the phase of structure factor $S(\vec{k}, \vec{r})$. Inheriting from the Ewald summation, the choice of the cell will not influence the predicted energy. So our reciprocal space potential keeps invariance of E(3) symmetry and choices of different cells. Also, the full computation process is differentiable and the forces on each atom can be computed with automatic differentiation.

During training, a three-step strategy is applied [45]. In the first stage, only the real space part will be used to fit the potential of the datasets. After the convergence is achieved, it comes to the second stage in which the real space NN will be frozen and the long-range part will be added to fit. The real space NN is not trained in this stage but helps to fit the potential with the long-range part. Along with the second convergence of the long-range NN, the third stage begins where the real space and the reciprocal space NN are employed to fit the potential together. With this three-step training strategy, we usually get a better potential than training the real part and reciprocal part together from the beginning as it works in the training of the neural network spin Hamiltonian [45]. This three-step strategy is noted as three-steps while the general training strategy of training together is noted as one-step in the following experiments.

***Method: Long-Range Descriptor in Reciprocal Space***. Further, we put forward a learnable reciprocal space descriptor to predict global crystal properties combined with local real space descriptors. Just like the way we handled long-range potentials, we build the long-range descriptor in reciprocal space based on the structure factor $|S(\vec{k},\vec{r})|$ and the length of $\vec{k}$ vector as $h_{long} = \sum_k \text{FCN}\left(|\vec{k}|, \frac{|S(\vec{k},\vec{r})|}{\Omega}\right)\left|\frac{S(\vec{k},\vec{r})}{\Omega}\right|^2$ with a cutoff of $k_{cutoff}$. Here, the charge $q_j$ in $S(\vec{k},\vec{r})$ is only related to the atomic number of the atoms and predicted by an embedding layer as represented as $q_j = \text{Embedding}(z_j)$. The E(3) symmetry invariance is kept as the long-range potential. Here we employ $\frac{|S(\vec{k},\vec{r})|}{\Omega}$ as the input of the neural network instead of $|S(\vec{k},\vec{r})|$ to achieve the invariance of choice of different cells. The number of atoms in a crystal differs in distinct cells resulting in different $|S(\vec{k},\vec{r})|$ as a summation of atoms but $\frac{|S(\vec{k},\vec{r})|}{\Omega}$ keeps the same with a division of the lattice volume. The reciprocal descriptors include the full atomic information and thus can represent the global periodic information of the crystal. The final full crystal descriptor including long-range information is concatenated with the local atomic crystal and the reciprocal crystal. With the final FCN layers and a pooling layer, the property of the crystal is predicted. A summation or average pooling is applied based on the property according to the pooling influence [46]. For extensive materials properties such as total energy, the full descriptor is first constructed for each atom and then a summation pooling is applied as $\text{Output} = \sum_i \text{FCN}(h_{short}^i \oplus h_{long})$ and and then go through the FCN layer. For intensive properties including band gap, the average pooling is taken on the short-range descriptor and then concatenated with the long-range descriptor as the full descriptor for the whole crystal. With a final FCN layer, the intensive property is predicted as $\text{Output} = \text{FCN}(\frac{1}{N_a}\sum_i h_{short}^i \oplus h_{long})$. The long-range potential can also be built with the long-range reciprocal descriptor as the extensive property prediction.

***Results:*** Firstly, the long-range descriptor is applied to the prediction of band gaps, an important intensive property, in the HOIP dataset [47]. It contains 1333 crystals of hybrid organic-inorganic perovskites. Dimenet++ [23] is employed as the CGNN in Figure 1. As for ablation experiments, four configurations are used to predict the band gap with the pure Dimenet++, the pure reciprocal descriptor, Dimenet++ with reciprocal potential, and Dimenet++ with the reciprocal descriptor. All of them are trained on the same training dataset and evaluated on the same testing dataset with results shown in Table 1. Dimenet++ with reciprocal descriptor achieves the best performance. Dimenet++ with reciprocal potential gives worse result than the pure Dimenet++ as expected because it's unreasonable to predict band gap with an addition of the real space and reciprocal space. The ablation results of the pure Dimenet++ and pure reciprocal descriptor shows that the reciprocal descriptor helps to encode the crystal information and predict the band gap while the reciprocal descriptor alone is not good enough for prediction. As shown in results, band gap is predicted better with the reciprocal descriptor.

| Methods | Test MAE (eV) of HOIP band gap |
|---|---|
| GIN [48] | 0.666 |
| CGCNN [6] | 0.170 |
| OGCNN [49] | 0.164 |
| CT [50] | 0.140 |
| Dimenet++ | 0.154 |
| Reciprocal descriptor | 0.385 |
| Dimenet++ with Reciprocal potential | 0.175 |
| Dimenet++ with Reciprocal descriptor | **0.129** |

Table 1. The test mean average error (MAE) of Dimenet++ with reciprocal descriptor in comparison to the supervised baselines and ablation configuration on the band gap of HOIP. The best results are shown in boldface.

Then, we focus on the long-range potential. A NaCl dataset with Coulomb interaction is built. The dataset is built on a 3×3×3 supercell of NaCl including 10,000 structures of 54 atoms with random perturbation on atoms and a fixed charge of +1e and -1e for Na and Cl respectively. 8,000 of them are used as training datasets with the left 2,000 as validation and test dataset. The energy label is calculated and classified as total energy and reciprocal energy with Ewald summation of Coulomb interaction implemented in pymatgen [51]. Dimenet++ is employed as the CGNN for real space potential. As for the reciprocal energy, the reciprocal space potential and Dimenet++ are applied to fit. During the experiments, we employ two configurations of pure Dimenet++ and Dimenet++ with reciprocal space potential for ablation comparison. The model is trained on the 3×3×3 supercell dataset and tested on 3×3×3 and 6×6×6 supercells with defects. Test results are shown in Table 2. The reciprocal space potential achieves good accuracy in the experiments and shows superiority on the complex 6×6×6 system with one point defect. Also, the 3-step strategy outperforms 1 step strategy in all experiments, yielding the importance of this training strategy.

| Method \ Data | Dimenet++ | Reciprocal space potential | Dimenet++ with reciprocal space potential (1 step) | Dimenet++ with reciprocal space potential (3 steps) |
|---|---|---|---|---|
| 3×3×3 Reciprocal energy | 3.18e-4 | **1.02e-4** | - | - |
| 3×3×3 Total energy | 10.5e-4 | - | 9.81e-4 | **5.86e-4** |
| 6×6×6 Total energy | **4.19e-3** | - | 5.32e-3 | 4.63e-3 |
| 6×6×6 with a defect Total energy | 1.83e-2 | | 1.87e-2 | **1.33e-2** |

Table 2. The test MAE per atoms of Dimenet++ with reciprocal space potential or reciprocal space potential in comparison to the pure Dimenet++ on the Coulomb energy of model NaCl dataset. The best results are shown in boldface.

The long-range interaction dominates in the defect system. Here we focus on Gallium Nitride the binary semiconductor with a direct band gap and $Ga_xN_y$ with different kinds of defects. A first principle dataset of $Ga_xN_y$ is built including 67208 different structures with corresponding energy and force. For the charge, the embedding layer with a neutral charge assumption is employed. We apply PaiNN [40] as the CGNN here because Dimenet++ as a line graph consumes much more memory and time to train such a large dataset. 80% of data are used to train with 10% for validating and 10% for tests. Test results are shown in Table 3. Compared with MAE, the root mean square error (RMSE) is a better metric that can reflect the fluctuations of the energy surface as the stability of the methods. As above experiments, the consideration of the reciprocal space energy and the 3-step strategy remarkably improves the performance and the stability of the potential. Less unfitted dusty points appear with the reciprocal space potential and almost vanished with the 3 steps strategy in Figure 3. It shows the importance of the reciprocal space potential in defect systems.

| Method / Label | PaiNN | PaiNN with reciprocal space potential (1 step) | PaiNN with reciprocal space potential (3 steps) |
|---|---|---|---|
| MAE Energy (meV) | 4.666 | 3.445 | **3.087** |
| RMSE Energy (meV) | 19.17 | 13.21 | **9.259** |
| MAE Force (meV/A) | 47.52 | 44.83 | **42.66** |
| RMSE Force (meV/A) | 414.6 | 250.9 | **249.5** |

Table 3. The test error of PaiNN with reciprocal space potential or reciprocal space potential in comparison to the pure PaiNN on the first-principle energy and force of the $Ga_xN_y$ dataset. The best results are shown in boldface.

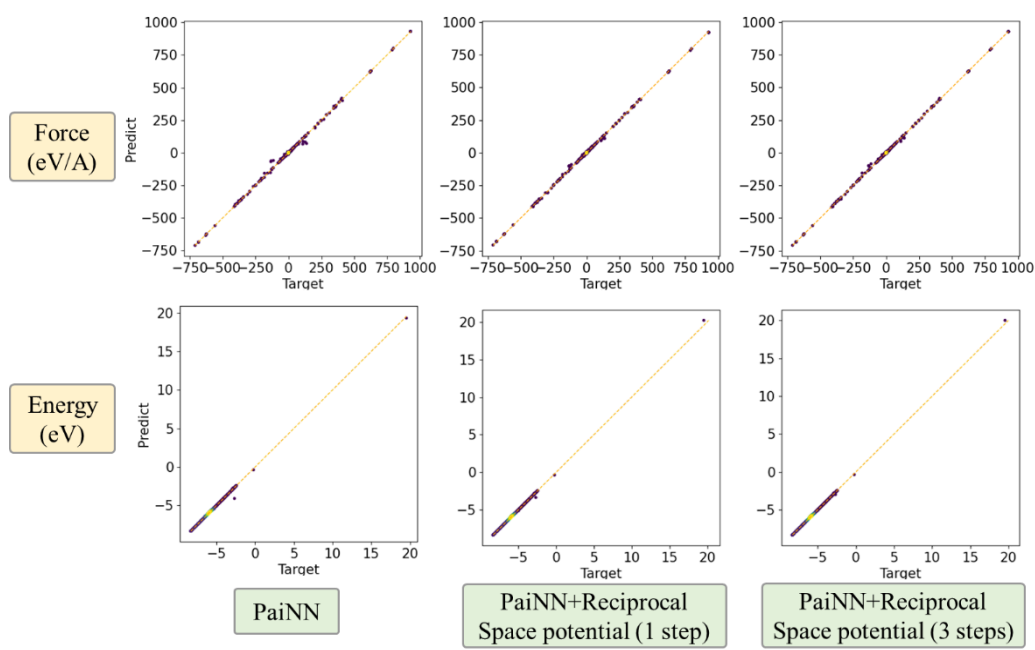

Figure 3. Comparison among energies and forces from DFT with different methods.

***Conclusion:*** In this article, we put forward the reciprocal space neural network for building long-range potential and descriptors. The reciprocal space descriptor with full atomic interaction and invariance helps to describe the crystal combined with local structure representation. Dimenet++ with the reciprocal space descriptor has achieved state-of-the-art accuracy on the band gap prediction. The reciprocal space potential is employed to describe the atomic long-range interactions in a neural-network style beyond the Coulomb interaction and successfully learn the model NaCl Coulomb interaction and help to build a more accurate and stable potential for the complex defect $Ga_xN_y$ system. With the reciprocal space information, we have enhanced the power of the crystal representation and the crystal atomic potential. It's promising in building potential models for defect systems and others in which long-range and full atomic interaction should be taken into account.

***Acknowledgments***. The work at Fudan is supported by NSFC (11825403, 11991061).

* Contributed equally to this work.

† hxiang@fudan.edu.cn

# Reference

[1] J. Behler and M. Parrinello, *Generalized Neural-Network Representation of High-Dimensional Potential-Energy Surfaces*, Phys. Rev. Lett. **98**, 146401 (2007).

[2] A. P. Bartók, M. C. Payne, R. Kondor, and G. Csányi, *Gaussian Approximation Potentials: The Accuracy of Quantum Mechanics, without the Electrons*, Phys. Rev. Lett. **104**, 136403 (2010).

[3] M. Rupp, A. Tkatchenko, K.-R. Müller, and O. A. von Lilienfeld, *Fast and Accurate Modeling of Molecular Atomization Energies with Machine Learning*, Phys. Rev. Lett. **108**, 058301 (2012).

[4] A. V. Shapeev, *Moment Tensor Potentials: A Class of Systematically Improvable Interatomic Potentials*, Multiscale Model. Simul. **14**, 1153 (2016).

[5] A. P. Thompson, L. P. Swiler, C. R. Trott, S. M. Foiles, and G. J. Tucker, *Spectral Neighbor Analysis Method for Automated Generation of Quantum-Accurate Interatomic Potentials*, J. Comput. Phys. **285**, 316 (2015).

[6] T. Xie and J. C. Grossman, *Crystal Graph Convolutional Neural Networks for an Accurate and Interpretable Prediction of Material Properties*, Phys. Rev. Lett. **120**, 145301 (2018).

[7] J. Behler, *Four Generations of High-Dimensional Neural Network Potentials*, Chem. Rev. **121**, 10037 (2021).

[8] T. B. Blank, S. D. Brown, A. W. Calhoun, and D. J. Doren, *Neural Network Models of Potential Energy Surfaces*, J. Chem. Phys. **103**, 4129 (1995).

[9] S. Gupta and D. Mukamel, *Relaxation Dynamics of Stochastic Long-Range Interacting Systems*, J. Stat. Mech. Theory Exp. **2010**, P08026 (2010).

[10] S. Gupta and D. Mukamel, *Slow Relaxation in Long-Range Interacting Systems with Stochastic Dynamics*, Phys. Rev. Lett. **105**, 040602 (2010).

[11] N. Artrith, T. Morawietz, and J. Behler, *High-Dimensional Neural-Network Potentials for Multicomponent Systems: Applications to Zinc Oxide*, Phys. Rev. B **83**, 153101 (2011).

[12] T. W. Ko, J. A. Finkler, S. Goedecker, and J. Behler, *A Fourth-Generation High-Dimensional Neural Network Potential with Accurate Electrostatics Including Non-Local Charge Transfer*, Nat. Commun. **12**, 1 (2021).

[13] A. Gao and R. C. Remsing, *Self-Consistent Determination of Long-Range Electrostatics in Neural Network Potentials*, Nat. Commun. **13**, 1 (2022).

[14] L. Zhang, H. Wang, M. C. Muniz, A. Z. Panagiotopoulos, R. Car, and W. E, *A Deep Potential Model with Long-Range Electrostatic Interactions*, J. Chem. Phys. **156**, 124107 (2022).

[15] E. R. Khajehpasha, J. A. Finkler, T. D. Kühne, and S. A. Ghasemi, *CENT2: Improved Charge Equilibration via Neural Network Technique*, Phys. Rev. B **105**, 144106 (2022).

[16] C. Ying, T. Cai, S. Luo, S. Zheng, G. Ke, D. He, Y. Shen, and T.-Y. Liu, *Do Transformers Really Perform Bad for Graph Representation?*, ArXiv210605234 Cs (2021).

[17] F. Wu, Q. Zhang, D. Radev, J. Cui, W. Zhang, H. Xing, N. Zhang, and H. Chen, *3D-Transformer: Molecular Representation with Transformer in 3D Space*, ArXiv211001191 Cs Q-Bio (2021).

[18] Y. Choukroun and L. Wolf, *Geometric Transformer for End-to-End Molecule Properties Prediction*, ArXiv211013721 Cs (2021).

[19] S. Batzner, A. Musaelian, L. Sun, M. Geiger, J. P. Mailoa, M. Kornbluth, N. Molinari, T. E. Smidt, and B. Kozinsky, *E(3)-Equivariant Graph Neural Networks for Data-Efficient and Accurate Interatomic Potentials*, Nat. Commun. **13**, 1 (2022).

[20] C. Chen, W. Ye, Y. Zuo, C. Zheng, and S. P. Ong, *Graph Networks as a Universal Machine Learning Framework for Molecules and Crystals*, Chem. Mater. **31**, 3564 (2019).

[21] J. Cheng, C. Zhang, and L. Dong, *A Geometric-Information-Enhanced Crystal Graph Network for Predicting Properties of Materials*, Commun. Mater. **2**, 1 (2021).

[22] K. T. Schütt, H. E. Sauceda, P.-J. Kindermans, A. Tkatchenko, and K.-R. Müller, *SchNet – A Deep Learning Architecture for Molecules and Materials*, J. Chem. Phys. **148**, 241722 (2018).

[23] J. Klicpera, S. Giri, J. T. Margraf, and S. Günnemann, *Fast and Uncertainty-Aware Directional Message Passing for Non-Equilibrium Molecules*, ArXiv201114115 Phys. (2020).

[24] F. Bloch, *Über die Quantenmechanik der Elektronen in Kristallgittern*, Z. Für Phys. **52**, 555 (1929).

[25] S. Li, Y. Liu, D. Chen, Y. Jiang, Z. Nie, and F. Pan, *Encoding the Atomic Structure for Machine Learning in Materials Science*, WIREs Comput. Mol. Sci. **12**, (2022).

[26] A. Ziletti, D. Kumar, M. Scheffler, and L. M. Ghiringhelli, *Insightful Classification of Crystal Structures Using Deep Learning*, Nat. Commun. **9**, 1 (2018).

[27] Z. Ren et al., *An Invertible Crystallographic Representation for General Inverse Design of Inorganic Crystals with Targeted Properties*, Matter **5**, 314 (2022).

[28] A. P. Bartók, R. Kondor, and G. Csányi, *On Representing Chemical Environments*, Phys. Rev. B **87**, 184115 (2013).

[29] O. Isayev, C. Oses, C. Toher, E. Gossett, S. Curtarolo, and A. Tropsha, *Universal Fragment*

*Descriptors for Predicting Properties of Inorganic Crystals*, Nat. Commun. **8**, 1 (2017).

[30] R. Drautz, *Atomic Cluster Expansion for Accurate and Transferable Interatomic Potentials*, Phys. Rev. B **99**, 014104 (2019).

[31] L. Himanen, M. O. J. Jäger, E. V. Morooka, F. Federici Canova, Y. S. Ranawat, D. Z. Gao, P. Rinke, and A. S. Foster, *DScribe: Library of Descriptors for Machine Learning in Materials Science*, Comput. Phys. Commun. **247**, 106949 (2020).

[32] K. Choudhary and B. DeCost, *Atomistic Line Graph Neural Network for Improved Materials Property Predictions*, Npj Comput. Mater. **7**, 1 (2021).

[33] J. S. Smith, O. Isayev, and A. E. Roitberg, *ANI-1: An Extensible Neural Network Potential with DFT Accuracy at Force Field Computational Cost*, Chem. Sci. **8**, 3192 (2017).

[34] J. Nigam, M. Willatt, and M. Ceriotti, *Equivariant Representations for Molecular Hamiltonians and N-Center Atomic-Scale Properties*, J. Chem. Phys. 5.0072784 (2021).

[35] M. Pinheiro, F. Ge, N. Ferré, P. O. Dral, and M. Barbatti, *Choosing the Right Molecular Machine Learning Potential*, Chem. Sci. **12**, 14396 (2021).

[36] I. Batatia, D. P. Kovács, G. N. C. Simm, C. Ortner, and G. Csányi, *MACE: Higher Order Equivariant Message Passing Neural Networks for Fast and Accurate Force Fields*, arXiv:2206.07697.

[37] J. Cheng, C. Zhang, and L. Dong, *A Geometric-Information-Enhanced Crystal Graph Network for Predicting Properties of Materials*, Commun. Mater. **2**, 1 (2021).

[38] J. Klicpera, F. Becker, and S. Günnemann, *GemNet: Universal Directional Graph Neural Networks for Molecules*, ArXiv210608903 Phys. Stat (2021).

[39] J. Klicpera, J. Groß, and S. Günnemann, *Directional Message Passing for Molecular Graphs*, ArXiv200303123 Phys. Stat (2020).

[40] K. T. Schütt, O. T. Unke, and M. Gastegger, *Equivariant Message Passing for the Prediction of Tensorial Properties and Molecular Spectra*, ArXiv210203150 Phys. (2021).

[41] D. Frenkel and B. Smit, *Chapter 12 - Long-Range Interactions*, in *Understanding Molecular Simulation (Second Edition)*, edited by D. Frenkel and B. Smit, Second Edition (Academic Press, San Diego, 2002), pp. 291–320.

[42] P. P. Ewald, *Die Berechnung Optischer Und Elektrostatischer Gitterpotentiale*, Ann. Phys. **369**, 253 (1921).

[43] M. Di Pierro, R. Elber, and B. Leimkuhler, *A Stochastic Algorithm for the Isobaric–Isothermal Ensemble with Ewald Summations for All Long Range Forces*, J. Chem. Theory Comput. **11**, 5624 (2015).

[44] D. E. Williams, *Accelerated Convergence of Crystal-Lattice Potential Sums*, Acta Crystallogr. Sect. A **27**, 452 (1971).

[45] H. Yu, C. Xu, X. Li, F. Lou, L. Bellaiche, Z. Hu, X. Gong, and H. Xiang, *Complex Spin Hamiltonian Represented by an Artificial Neural Network*, Phys. Rev. B **105**, 174422 (2022).

[46] S. Gong, T. Xie, Y. Shao-Horn, R. Gomez-Bombarelli, and J. C. Grossman, *Examining Graph Neural Networks for Crystal Structures: Limitation on Capturing Periodicity*, arXiv:2208.05039.

[47] C. Kim, T. D. Huan, S. Krishnan, and R. Ramprasad, *A Hybrid Organic-Inorganic Perovskite Dataset*, Sci. Data **4**, 1 (2017).

[48] W. Hu, B. Liu, J. Gomes, M. Zitnik, P. Liang, V. Pande, and J. Leskovec, *Strategies for Pre-Training Graph Neural Networks*, Int. Conf. Learn. Represent. ICLR (2019).

[49] M. Karamad, R. Magar, Y. Shi, S. Siahrostami, I. D. Gates, and A. Barati Farimani, *Orbital Graph Convolutional Neural Network for Material Property Prediction*, Phys. Rev. Mater. **4**, 093801 (2020).

[50] R. Magar, Y. Wang, and A. Barati Farimani, *Crystal Twins: Self-Supervised Learning for Crystalline Material Property Prediction*, Npj Comput. Mater. **8**, 1 (2022).

[51] S. P. Ong, W. D. Richards, A. Jain, G. Hautier, M. Kocher, S. Cholia, D. Gunter, V. L. Chevrier, K. A. Persson, and G. Ceder, *Python Materials Genomics (Pymatgen): A Robust, Open-Source Python Library for Materials Analysis*, Comput. Mater. Sci. **68**, 314 (2013).